\documentclass{aa}

\usepackage{graphicx}
\usepackage{txfonts,textcomp}
\usepackage[normalem]{ulem} % for \sout

\newcommand\rurl[1]{%
  \href{https://#1}{\nolinkurl{#1}}%
}
\makeatletter
\renewcommand*\aa@pageof{, page \thepage{} of \pageref*{LastPage}}

\usepackage{natbib,twoopt}
\usepackage[breaklinks]{hyperref}      %% to avoid \citeads line fills, add "draft" 
\bibpunct{(}{)}{;}{a}{}{,}             %% natbib format for A&A and ApJ

\hypersetup{
  colorlinks,
  citecolor=blue,
  linkcolor=blue,
  urlcolor=blue,
}
\makeatletter
  \newcommandtwoopt{\citeads}[3][][]{\href{http://adsabs.harvard.edu/abs/#3}%
    {\def\hyper@linkstart##1##2{}%
     \let\hyper@linkend\@empty\citealp[#1][#2]{#3}}}
  \newcommandtwoopt{\citepads}[3][][]{\href{http://adsabs.harvard.edu/abs/#3}%
    {\def\hyper@linkstart##1##2{}%
     \let\hyper@linkend\@empty\citep[#1][#2]{#3}}}
  \newcommandtwoopt{\citetads}[3][][]{\href{http://adsabs.harvard.edu/abs/#3}%
    {\def\hyper@linkstart##1##2{}%
     \let\hyper@linkend\@empty\citet[#1][#2]{#3}}}
  \newcommandtwoopt{\citeyearads}[3][][]%
    {\href{http://adsabs.harvard.edu/abs/#3}
    {\def\hyper@linkstart##1##2{}%
     \let\hyper@linkend\@empty\citeyear[#1][#2]{#3}}}
\makeatother

\begin{document}

   \title{XRBcats: Galactic low-mass X-ray binary catalogue
   \thanks{The catalogue is available at the CDS via anonymous ftp to \href{https://cdsarc.cds.unistra.fr}{cdsarc.cds.unistra.fr (130.79.128.5)} or via \url{https://cdsarc.cds.unistra.fr/viz-bin/cat/J/A+A/675/A199}, and the catalogue table, along with the corresponding finding charts and other useful information is also publicly accessible at \url{http://astro.uni-tuebingen.de/~xrbcat}.}
   }

   \author{A. Avakyan\inst{1}\thanks{E-mail: artur.avakyan@astro.uni-tuebingen.de},
    M. Neumann\inst{1},
   A. Zainab\inst{2},
          V. Doroshenko\inst{1},
          J. Wilms\inst{2}
        \and 
        A. Santangelo\inst{1}.
          }
\authorrunning{author}
          
   \institute{\inst{1}Universit{\"a}t T{\"u}bingen, Institut f{\"u}r Astronomie und Astrophysik T{\"u}bingen, Sand 1, 72076 T{\"u}bingen, Germany\\
            \inst{2}Dr. Karl-Remeis Sternwarte and Erlangen Centre for Astroparticle Physics, Friedrich-Alexander Universit{\"a}t Erlangen-N{\"u}rnberg, Sternwartstr. 7, 96049 Bamberg, Germany
            }

   \date{Received 28 March 2023 / Accepted 26 May 2023}

  \abstract{
    We present a new catalogue of low-mass X-ray binaries (LMXBs) in
    the Galaxy. The catalogue contains source names, coordinates, source types, fluxes, distances, system parameters, and other characteristic properties of 349 LMXBs, including systems that have been newly discovered or reclassified since the most recently reported LMXB catalogues. 
    The aim of this catalogue is to provide a list of all currently known Galactic objects identified as LMXBs with some basic information on each system (including X-ray and optical/IR properties where possible). Literature published before May 2023 has been taken into account where possible when compiling this information. References for all reported properties as well as object-finding charts in several energy bands are provided as part of the catalogue. We plan to update the catalogue regularly, in particular to reflect new objects discovered in the ongoing large-scale surveys such as \textit{Gaia} and eROSITA.}

   \maketitle
%
%-------------------------------------------------------------------
\section{Introduction}

X-ray binaries (XRBs) are among the brightest objects in the sky in the X-rays. They represent the endpoints of the stellar evolution of massive stars. Understanding the physical processes defining the observational appearance of individual objects and the properties of the XRB population as a whole is therefore essential for understanding massive star evolution and the evolution of the Galaxy in general. To this end, it is important to maintain an up-to-date census of known XRBs and their properties, especially in the era of the current generation of large-scale surveys such as the extended ROentgen Survey Imaging Telescope Array~\citep[eROSITA,][]{Merloni12, eROSITA} on board SRG, \textit{Gaia}~\citep{Gaia_miss,Gaiadr3} and the Wide-field Infrared Survey Explorer \citep[WISE,][]{WISE, WISE_cat}, which are able to unveil ever fainter objects.

However, the identification of XRBs among millions of other objects requires up-to-date knowledge of the properties and locations of already known sources, and the large-scale surveys mentioned above are providing vast amounts of
new observational data for these objects. The compilation of such an updated database is the primary goal of this work, although many other uses can of course also be envisaged.

The main source of high-energy radiation in XRBs is the accretion of matter onto either a neutron star (NS) or a black hole (BH) from a companion star. The observed properties of XRBs are largely defined by the mass-transfer mechanism powering the accretion, which can occur either directly from the wind of a secondary or via Roche lobe overflow (RLOF), which primarily depends on the mass ratio between the compact object and the optical counterpart~($M_{\rm x}/M_{\rm opt}$). Based on this ratio, XRBs are subdivided into two large groups: high- and low-mass X-ray binaries (HMXBs and LMXBs, respectively). For objects with $M_{\rm x}>M_{\rm opt}; M_{\rm opt}\lesssim1M_\odot$, the mass is usually transferred to the compact object via RLOF. Such objects are classified as LMXBs. Those with $M_{\rm x} < M_{\rm opt}; M_{\rm opt}\gtrsim 5 M_\odot$ typically accrete directly from the stellar wind and are classified as HMXBs. The rather dramatic differences in the observed properties of LMXBs and HMXBs warrant the treatment of each subclass  separately. Here, we focus  on the properties of LMXBs. In turn, updated catalogues of HMXBs are presented by \citealt{Marvin23} and \citet{Fotin}\footnote{\url{https://vizier.cds.unistra.fr/viz-bin/VizieR?-source=J/A+A/671/A149}}$^{,}$\footnote{\url{https://binary-revolution.github.io/HMXBwebcat}}. 

As follows from the definition of an LMXB, the donor is (in most cases) a late-spectral-type star filling its Roche lobe. However, A-type stars, F-G-type subgiants, or even white dwarfs (WDs) can also act as donors in LMXBs.
The optical properties of LMXBs can also be affected by emission from the accretion disc around the compact object, where the disc can be heated by itself or illuminated by X-ray emission from the compact object. Nevertheless, LMXBs are generally intrinsically faint objects in the optical and IR bands. The main way to analyse features of known LMXBs in detail or to detect new ones is to observe them during outbursts \citep{Las16}.
New sources continue to be discovered  in this way, and the population of known LMXBs is ever growing.

The latest catalogue of Galactic LMXBs, which was published by \citet{Liu07}, is now nearly 16 years old.
Since that time, many new transient and persistent objects have been discovered. Some LMXB candidates have been reclassified, while other sources have gained LMXB status. Moreover, missions such as the X-ray Multi-Mirror Mission~\citep[\textit{XMM-Newton},][]{XMM_mission}, the INTErnational Gamma-Ray Astrophysics Laboratory \citep[INTEGRAL,][]{Integral1}, and \textit{Gaia} have resulted in a quantitative and qualitative improvement in our knowledge of XRB properties. In this context, we have compiled a revised catalogue of the Galactic LMXBs that incorporates multi-wavelength information and encompasses the inclusion of new sources identified since the publications of \citet{Liu07} and  \citealt{RitterKolb04} (Liu07 and RK03 hereinafter). Our final catalogue contains 349 LMXBs and candidates, of which 204 were contained in the two catalogues mentioned above (the content of Liu07 and RK03 overlaps but neither includes the other completely). That is, the new catalogue presented here represents a more than 71\% increase in volume with respect to the two most extensive catalogues of the past. We aim to compile a comprehensive and up-to-date database containing information such as optical/IR magnitudes, X-ray fluxes, distances, and so on, for the current catalogue, which will hopefully help to facilitate the identification of new LMXBs in current and future X-ray surveys, as well as population studies.

The structure of the paper is as follows: in Sect.~\ref{sec:sample}, we describe how the LMXB sample was compiled and list data sources used to provide extra information on known objects. In the same section, we list relevant X-ray and optical properties, and characterise methods used for identification of known and candidate optical/IR counterparts in large-scale surveys. In Sect.~\ref{sec:cross}, a description of the table fields and finding charts is provided, as well as a comparison of our catalogue with works of RK03 and Liu07. Finally, we summarise our results in Sect.~\ref{conc}. One can also find a list of identified archive optical/IR counterparts for some LMXBs based on the literature in Appendix~\ref{append1}. A description of the columns of the catalogue table is presented in Appendix~\ref{append2}.

\section{Definition of the sample and data sources}\label{sec:sample}

The bulk of the objects in the catalogue consists of LMXBs reported by the catalogues of RK03 and Liu07, and so we started our compilation of the sample by merging these two catalogues. During this process, a number of duplicate objects both within and across the two catalogues were identified and removed from the sample. In addition, several sources in RK03 and Liu07 were found to be extragalactic (in some cases, this possibility was already indicated in the original catalogues and sometimes reported in the literature afterwards). 
Such objects were also excluded from our final list. 
The same applies to the satellite galaxies of the Milky Way. To this end, the two LMXBs known to be located in the Large Magellanic Cloud (LMC), namely, LMC~X$-$2 and RX\,J0532.7$-$6926 (which were included in both RK03 and Liu07) are not listed in our catalogue.
All objects that were removed, as well as other problematic cases, are listed and described in detail in Sect.~\ref{sec:except}. 

In addition to Liu07 and RK03, we also considered all Galactic objects classified as possible LMXBs in the SIMBAD\footnote{\url{http://simbad.cds.unistra.fr/simbad}} and VizieR\footnote{\url{https://vizier.cds.unistra.fr/viz-bin/VizieR}} databases hosted by the Centre de Donn\'{e}es astronomiques de Strasbourg (CDS) databases. We include these objects in the catalogue, giving appropriate references for classification. We conducted a thorough literature search (including Astronomer's Telegrams\footnote{\url{https://astronomerstelegram.org}}) to identify any newly discovered LMXBs.
Most of the new LMXBs reported in the literature between 2006 and 2020 were discovered by the INTEGRAL \citep{2003A&A...411L...1W}, the Monitor of All-sky X-ray Image~\citep[MAXI,][]{MAXI}, \textit{Swift} \citep{SWIFT}, or \textit{Gaia} \citep{Gaia_miss} missions, as summarised by \citet{Bahramian_lmxb}. The latter publication can therefore also be considered as another major data source used in this work. In addition, the catalogue of Galactic LMXBs detected by INTEGRAL~\citep{Sazonov_lmxb} also proved to be a valuable resource for expanding our list of LMXBs. Finally, we used the recently published catalogue of ultracompact X-ray binaries (UCXBs) made by \citealt{UltraCompCAT}\footnote{\url{https://research.iac.es/proyecto/compactos/UltraCompCAT}} to cross-check the completeness of our sample in the context of UCXBs (no missing sources were found).
It should be noted that although efforts have been made to find all LMXBs identified to date, it is possible that some may have been missed. Therefore, we kindly ask readers to report any omissions found, so that those can be included in updated versions of the catalogue.

We note that systems in which a WD acts as an accretor are not generally considered as XRBs for historical reasons (despite the fact that they also emit in the X-ray band), and so those systems (namely cataclysmic variables or CVs) are not included in the current catalogue. 
We also do not include rotation-powered pulsars, even if they are members of binary systems and emit in X-rays, unless the observed emission is accretion powered; that is, `spider'-type pulsars including Black Widows~(BWs) and Redbacks~\citep[RBs; see e.g.][]{BW_Chen, BW_Roberts} are omitted unless they are also classified as transitional millisecond pulsars \citep[tMSP; see e.g.][]{2009Sci...324.1411A, 2013Natur.501..517P, 2023MNRAS.520.3416K}. All tMSPs and their candidates included in the catalogue are listed in  Table~\ref{tabl:tMSPcands} in Sect.~\ref{sec:except}.

The mass ranges quoted in Sect. 1 for XRBs are of course approximate, and there are some boundary cases, such as Her~X$-$1, where accretion to a NS occurs via the RLOF of a B3 type $\sim2.2 M_{\rm \odot}$ donor \citep{Her, Her1}. This system is most often classified in the literature as an intermediate mass X-ray binary~(IMXB) rather than as an LMXB because the NS appears as an X-ray pulsar and is therefore strongly magnetised, which is not typical for NS LMXBs. As the total number of IMXBs is relatively small, and  despite the additional fact that they `bridge' the HMXB and LMXB classes, they are evolutionarily closer to LMXBs as binary systems, and so we opt to include them as a part of the current LMXB catalogue. This decision is also clearly communicated in the publication describing the HMXB catalogue released in parallel to the current work \citep{Marvin23} in order to avoid any confusion or source duplication, except for one specific case (Cir~X$-$1), which is described in Sect.~\ref{sec:except}.  

    \begin{figure}
      \centering
    \includegraphics[width=\columnwidth]{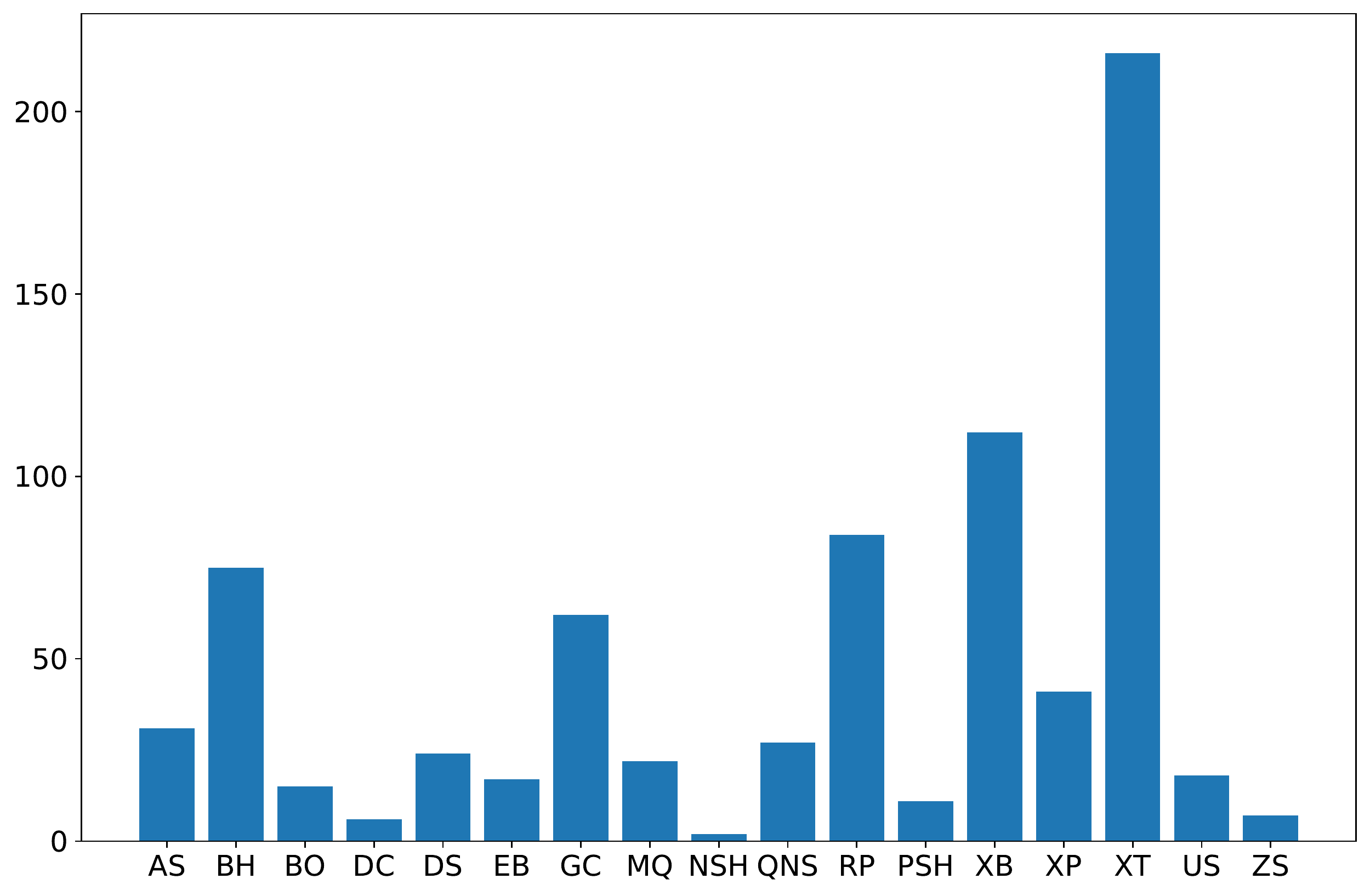}
    \caption{Population of LMXBs of each type in the Galaxy.} 
    \label{fig:hist}
    \end{figure}

\subsection{Main properties and features}\label{sec:xtypes}
In order to reflect the huge variety of LMXB properties, we incorporate all accessible and relevant information concerning the donor star and compact object, including data on their properties, variability, and other pertinent details.
Besides the quantitative properties (such as positions, fluxes in various bands, orbital or spin periods, and so on), some of these properties or features can be incorporated as a set of flags defined to characterise several LMXB subclasses, as was already done by Liu07 and RK03. Here, we extend the classification adopted by these authors to account for extra features added to the current catalogue. In order to provide a broad overview of the Galactic LMXB properties, we also show the frequency of the occurrence of the individual type flags in graphical format in Fig.~\ref{fig:hist}. It is important to note that these flags are not intended to serve as a basis for systematic classification, but rather to reflect the characteristics of individual systems to ease the selection of objects exhibiting specific observational features. Each source in the catalogue may therefore have one or more of the following flags, which are listed alphabetically:

\begin{enumerate}\label{flags}
\item AS: Atoll X-ray binary hosting a weakly magnetised NS as a main component. The spectrum is soft and no significant pulsations are present. The majority of the points on the colour--colour diagrams (`hard' versus `soft') from atoll sources usually form a band at constant hard colour.
    
\item BH: System contains a black hole candidate.
    
\item BO: X-ray burster (see XB definition below) with coherent burst oscillations at the NS spin period.

\item DC: System with an accretion disc corona.
    
\item DS: `Dipping' source. System shows periodic but irregular dips in the X-ray intensity, which are generally connected with the partial obscuration of the NS by a thickened region of the accretion disc.
    
\item  EB: Eclipsing or partially eclipsing binary system.
    
\item GC: Source within a globular cluster.
    
\item MQ: Microquasar, a source with reported evidence for relativistic jets.
   
\item NSH: Negative (nodal) superhumps are present in the system. These negative superhumps are periodic optical signals, the periods of which are smaller than the orbital period of the binary.

\item QNS: Quiescent NS.
   
\item RP: Compact object acts as a radio pulsar. 
   
\item PSH: Positive superhumps (permanent or transient) present in the system. The same as the NSH type (see above), but the time period of the signal is larger than the orbital one. 
   
\item XB: X-ray burst source. This type collects X-ray binary systems with a NS, which exhibit thermonuclear (type-I) X-ray bursts.

\item XP: Compact object in the binary acts as an X-ray pulsar. 

\item  XT: Transient X-ray source. System shows dramatic changes in luminosity (mostly in the X-ray band). Energy is produced by means of a non-stationary accretion process.

\item US: Ultrasoft X-ray spectrum. In some catalogues, this type is denoted `super soft'. These binaries also include some `extreme ultrasoft' (EUS) sources.

\item ZS: Z-type source. This type of X-ray binary is similar to the atoll type (also contains weakly magnetised NS), but the plot on the colour--colour diagram is different (Z-letter shape). 
\end{enumerate}

We note that the `XB' (X-ray burster) flag --- originally introduced by RK03 --- in our catalogue specifically refers to thermonuclear (type-I) bursts only. Type-II X-ray bursts likely caused by instabilities in the accretion flow are relatively rare, and to the best of our knowledge have only been detected from two LMXBs: MXB~1730$-$335~\citep[`rapid burster';][]{2015MNRAS.449..268B} and GRO~J1744$-$28 \citep[`bursting pulsar';][]{2018MNRAS.481.2273C}, and one HMXB: SMC~X$-$1~\citep{2018RAA....18..148R}, and are therefore not flagged separately. The MXB~1730$-$335 is known to exhibit bursts of both types, and is therefore still flagged.
On the other hand, type-I bursts have
been suggested to be present but have never been unambiguously detected from the bursting pulsar~\citep[]{1996ApJ...469L..25G, 1996astro.ph..4089L, 2015MNRAS.452.2490D}, and so this source is not flagged.

\subsection{X-ray properties}

Considering that X-ray emission dominates the bolometric luminosity of XRBs, one of their main characteristics is their X-ray flux and spectrum. In our catalogue, we therefore tried to compile information on X-ray fluxes in soft and hard X-ray bands for all LMXBs  in a uniform way. For the soft X-ray band, we report fluxes observed by \textit{XMM-Newton}\footnote{\url{https://heasarc.gsfc.nasa.gov/W3Browse/xmm-newton/xmmssc.html}}~\citep{Schartel2022,XMM}, the
\textit{Chandra} X-ray Observatory\footnote{\url{https://vizier.cds.unistra.fr/viz-bin/VizieR-3?-source=IX/57/csc2master}}~\citep[CXO,][]{CXO_miss, CXO}, and \textit{Swift}/XRT\footnote{\url{https://vizier.cds.unistra.fr/viz-bin/VizieR-3?-source=IX/58/2sxps}}~\citep{2005SSRv..120..165B,XRT} throughout their lifetime. In particular, we report the 0.2--12\,keV energy band flux (equivalent to the EPIC\_8 band used in \textit{XMM-Newton} catalogues) and/or the X-ray flux in similar
energy bands of CXO and \textit{Swift}/XRT, that is, the 0.3--10\,keV observed flux for \textit{Swift}/XRT, and either the broad ACIS band (0.5--7.0\,keV) or the wide HRC band (0.1--10.0\,keV) for CXO.
We note that although the difference in energy bands for various instruments might result in a factor of ${\sim}1.5$ difference for a given source, here we only report the range of fluxes detected over a long period. This range is therefore dominated by intrinsic variability and differences in source spectra, and so the difference in energy range between the various instruments is not really relevant  to our purposes. 
In the hard X-ray band, fluxes from INTEGRAL\footnote{\url{https://vizier.cds.unistra.fr/viz-bin/VizieR-3?-source=J/A\%2bA/545/A27}}~\citep{INTEGRAL} and \textit{Swift}/BAT\footnote{\url{https://heasarc.gsfc.nasa.gov/W3Browse/swift/swbat105m.html}}~\citep{BAT1, BAT} are reported. Here, we report fluxes in the energy range of 14-145\,keV for \textit{Swift}/BAT and in the energy range of 17--60\,keV for INTEGRAL. We note that these bands are largely equivalent as the flux above 60\,keV is comparatively small for most sources, and the variability argument still applies. 

Another key property provided by X-ray observations is the localisation that is essential to identify or assess the reliability of the identified optical counterparts. We therefore also include information on the most accurate X-ray position available either from the literature or directly from soft (\textit{Chandra}, \textit{XMM-Newton} or \textit{Swift}/XRT) and hard (INTEGRAL, \textit{Swift}/BAT) catalogues as described in the following section.

\subsection{Astrometry and identification of the optical counterparts}\label{sec:counter}

Considering that the properties of XRBs are largely defined by the properties of the donor star, the identification of the optical counterparts and the characterisation of their multi-wavelength properties are essential. Some of the LMXBs in the sample already have robustly identified optical counterparts; however, the optical positions (as well as other optical data) reported in Liu07 and RK03 are in many cases outdated, and lack accuracy by current standards. We therefore attempted to improve position accuracy using the latest astrometry data from \textit{Gaia}~DR2\footnote{\url{https://vizier.cds.unistra.fr/viz-bin/VizieR?-source=I/345}}, eDR3\footnote{\url{https://vizier.cds.unistra.fr/viz-bin/VizieR-3?-source=I/350/gaiaedr3}}, DR3\footnote{\url{https://vizier.cds.unistra.fr/viz-bin/VizieR-3?-source=I/355}}~\citep{Gaiadr2, Gaiaedr3, Gaiadr3}, Two Micron All Sky Survey\footnote{\url{https://vizier.cds.unistra.fr/viz-bin/VizieR?-source=II/246}}~\citep[2MASS,][]{2MASS_miss, 2MASS, 2MASS_1}
and the mid-IR CatWISE2020 catalogue\footnote{\url{https://vizier.cds.unistra.fr/viz-bin/VizieR?-source=II/365}}~\citep{CW2020}.

We started the process of identifying counterparts in these catalogues by verifying whether the SIMBAD database already had any pre-existing matches with \textit{Gaia}~DR3 (eDR3, DR2), 2MASS, and/or WISE and checking their correctness manually. It should be noted that in some cases, the associations provided by SIMBAD were indeed incorrect, and we list such cases in Sect.~\ref{sec:except}. In cases where no SIMBAD optical/IR identifiers were provided, we attempted to find counterparts reported in the literature. In some cases where precise optical, IR, or radio coordinates were available, these coordinates were used to find the corresponding counterparts in \textit{Gaia}~DR3, 2MASS, and CatWISE2020.
All solid cross-matches and relevant data were added to the catalogue. In particular, identifications by \citet{Arnason21} proved to be useful, and so these are considered as the main references for the cases marked in  Table~\ref{tabl:opt} in Appendix~\ref{append1}. This table lists all sources for which optical/IR counterparts are available, except those correctly listed in SIMBAD.

As mentioned above, some LMXBs have been classified as such only based on their X-ray properties, and have no uniquely identified optical counterparts in the literature. In most cases, these are relatively poorly studied sources (sometimes only detected once with X-ray monitors during an outburst). However, in some cases, accurate X-ray positions obtained at a later time are available. For those objects, we used CXO, \textit{XMM-Newton}, or \textit{Swift}/XRT positions (in order of preference). The search radius was set to match the full positional uncertainty (i.e. statistical plus systematic) reported in the respective X-ray catalogue. The final identification of plausible optical counterparts was done manually by inspection of the finding charts also released as part of the current catalogues and by searching for tentative counterparts in the literature based on their properties. In all cases where it was possible to identify a plausible counterpart, we list it as such; however, the optical position is only assigned as primary if a unique counterpart is identified. 

\subsection{Additional information}

For all optical counterparts identified as described above, we include information on IR and/or optical magnitudes wherever possible. In particular, we include \textit{Gaia} G-band or V-band magnitudes in the optical band, near-IR photometry either from 2MASS or Visible and Infrared Survey Telescope for Astronomy \citep[VISTA,][]{Vista, Vista1} Variables in the Via Lactea (VVV) DR4 catalogue \citep[VVV\footnote{\url{http://vvvsurvey.org}},][]{VVV} surveys, or from the literature. Finally, we also include magnitudes in the W1  and W2 bands reported in the CatWISE2020, AllWISE, or WISE (in that order of preference) catalogues~\citep{CW2020, WISE_cat, WISE}. 
In all cases, we provide the references to the sources where the V-band, G-band, JHK, and WISE magnitudes were taken from and show them in `Vmag\_Ref', `Gmag\_Ref', `JHK\_Ref', and `WISE\_Ref' columns, respectively. It should be noted that we mainly tried to include the optical/IR magnitudes during the quiescent state of the corresponding binary (to represent the LMXB population in the state in which it is found most of the time). However, in some cases, `quiescent' magnitudes were not available, and so we list magnitudes during the outburst, if any.

Besides more accurate positions and photometry, we also include some additional information unavailable at the time of publication of Liu07 and RK03. 
For all LMXBs for which a \textit{Gaia} counterpart was ultimately identified, we provide distance estimates from \textit{Gaia}~DR3, those based on \textit{Gaia}~eDR3\footnote{\url{https://vizier.cds.unistra.fr/viz-bin/VizieR-3?-source=I/352}}~\citep{Gaiaedr3dist} or \textit{Gaia}~DR2\footnote{\url{https://vizier.cds.unistra.fr/viz-bin/VizieR-3?-source=I/347}}~\citep{Gaiadr2dist}, and  estimations from StarHorse1\footnote{\url{https://vizier.cds.unistra.fr/viz-bin/VizieR-3?-source=I/349/starhorse}}/2\footnote{\url{https://vizier.cds.unistra.fr/viz-bin/VizieR-3?-source=I/354/starhorse2021}}\citep{StarHorse19, StarHorse21}. In addition, we manually searched the literature for any distance estimations, especially in cases where \textit{Gaia}/StarHorse data were unavailable.

We were also interested in the spectral type of the optical counterpart, and we therefore attempted to include this information based on either \textit{Gaia}~DR3, SIMBAD, or the literature search. We note that the list of X-ray pulsars maintained by Mauro Orlandini\footnote{\url{http://www.iasfbo.inaf.it/~mauro/pulsar_list.html}} was found to be a useful resource in this regard (and also for reported spin and orbital periods) and we would like to acknowledge it here. For sources where the reported spectral type and corresponding reference were published before 2007, the spectral information in RK03 or Liu07 was used. 

Finally, for convenience, we also provide the X-ray absorption neutral hydrogen column density ($N_{\rm H}$) estimates, which  might be relevant to estimate some additional LXMB properties such as X-ray to optical luminosity ratios in order to facilitate searches for new LMXBs with similar properties. 
We report values based either on \textit{Swift}/XRT data~\citep{XRT} or on results reported in the literature (all references are listed). 
In cases where one of the two ($N_{\rm H}$ or $A_{\rm G}$) was unknown, we calculated the other one through the $N_{\rm H}(A_{\rm G})$ relation reported by \citet{Ag_NH}.

\section{Catalogue content and quality assurance}\label{sec:cross}
\subsection{Description of the fields}\label{sec:fields}

There are 349 entries in the catalogue, each of which corresponds to a single Galactic source. The objects are sorted by right ascension (the second column) in increasing order. There are a total of 62 columns listing various parameters and corresponding references. The content and format of individual columns is described in detail in Table~\ref{tabl:cols} in Appendix~\ref{append2}.

\subsection{Finding charts and problematic cases}\label{sec:except}

We provide finding charts for all sources in our catalogue, which includes up to six different images, covering bands from mid-IR to hard X-rays. Corresponding charts can be found on a dedicated website\footnote{\url{http://astro.uni-tuebingen.de/~xrbcat}} or obtained via CDS system\footnote{\url{https://cdsarc.cds.unistra.fr/viz-bin/cat/J/A+A/675/A199}}. The top row of each finding chart includes images from the VVV DR4 catalogue\footnote{\url{http://alasky.cds.unistra.fr/VISTA/VVV_DR4/VISTA-VVV-DR4-J}}~\citep{VVV} or 2MASS\footnote{\url{http://alasky.cds.unistra.fr/2MASS/J}} (left, in order of preference), an image of unWISE\footnote{\url{http://alasky.cds.unistra.fr/unWISE/W1}}~\citep{2019ApJS..240...30S} (middle), and an RGB-image of either \textit{Chandra}\footnote{\url{https://cdaftp.cfa.harvard.edu/cxc-hips}}, \textit{XMM-Newton}\footnote{\url{http://skies.esac.esa.int/XMM-Newton/EPIC-RGB}}, or the Roentgensatellit (ROSAT\footnote{\url{http://alasky.cds.unistra.fr/RASS}}, \citealt{ROSAT_miss, ROSAT, ROSAT99}) (right, in order of preference).

The bottom row includes the soft X-ray image of \textit{Swift}/XRT (SwiftXRTInt in astroquery.skyview) in the left corner, and hard X-ray images from \textit{Swift}/BAT\footnote{\url{http://cade.irap.omp.eu/documents/Ancillary/4Aladin/BAT_14_20}}  and INTEGRAL\footnote{\url{http://cade.irap.omp.eu/documents/Ancillary/4Aladin/INTEGRAL_17_60}}  in the middle and right corners, respectively. The field of view for VVV, 2MASS, unWISE, \textit{XMM-Newton}, and \textit{Chandra} images was set at 1 arcmin, while for \textit{Swift}/XRT and ROSAT it was set at 5 arcmin and 15 arcmin, respectively. The field of view for \textit{Swift}/BAT and INTEGRAL was chosen to be $10^\circ$, with the size of the region chosen based on the field of view and angular resolution of the given instrument.

Each panel of the finding chart includes the coordinates and uncertainties of all detected sources within the respective regions, with position uncertainties represented by error circles. A red cross indicates the position listed in SIMBAD for the source, while a dark-blue diamond represents the coordinates described in the literature~(Liu07 or RK03). The position used for this catalogue is indicated with an orange star. In the case of soft X-ray instruments, observation positions are indicated by error circles to prevent overcrowding.
A golden circle indicates the \textit{Chandra} position, a red circle indicates the \textit{XMM-Newton} position, a green circle indicates the \textit{Swift}/XRT position, and a navy-blue circle indicates the ROSAT position. \textit{Swift}/BAT and INTEGRAL are indicated as a deep-pink pentagon and a lime-green triangle, respectively. In the optical band, we indicate CatWISE2020 with an orange `X', 2MASS with a cyan `+', and \textit{Gaia}~DR3 data with a purple square.
From a technical perspective, we used the Hierarchical Progressive Surveys (HiPS) data from \citealt{2015A&A...578A.114F}, using SkyView Query only for \textit{Swift}/XRT to obtain images. Both services were queried using  \texttt{astroquery.hips2fits}\footnote{\url{https://astroquery.readthedocs.io/en/latest/hips2fits/hips2fits.html}}, and \texttt{astroquery.skyview}\footnote{\url{https://astroquery.readthedocs.io/en/latest/skyview/skyview.html}} modules.

\begin{figure*}
    \includegraphics[width=\textwidth]{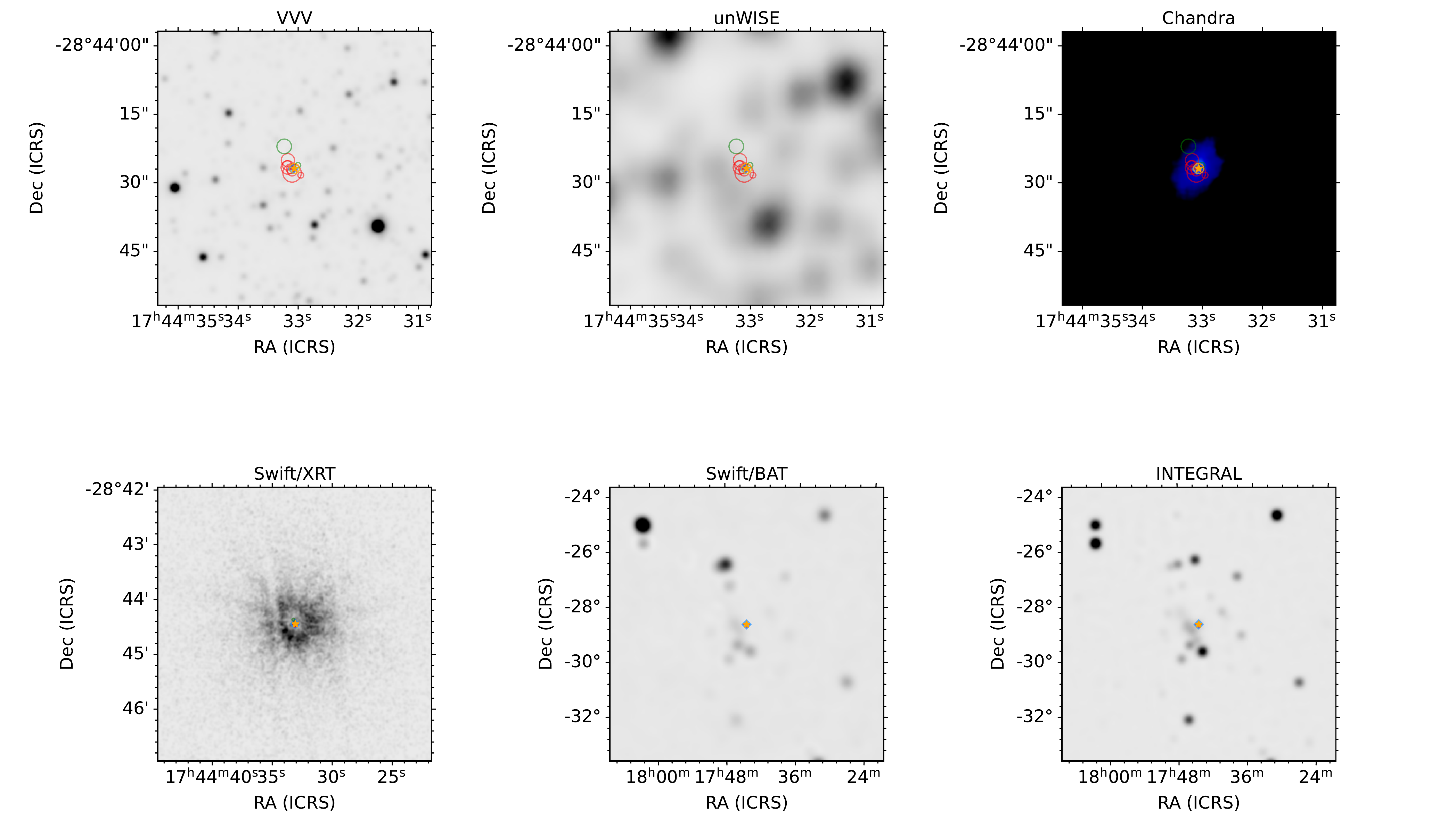}
    \caption{Finding chart of GRO\,J1744$-$28. The finding charts display markers and error circles representing observations made by different instruments. The error circles are used for soft X-ray observations to prevent overcrowding, with a yellow circle indicating \textit{Chandra} and a green circle representing \textit{Swift}/XRT. The red cross and the blue diamond indicate the positions reported by SIMBAD and that found in the literature (Liu07 or RK03), respectively. An orange star denotes the position of the LMXB used in this catalogue.}
    \label{fig:finding}
  \end{figure*}

As an example, the finding chart of GRO\,J1744$-$28 can be seen in Fig.~\ref{fig:finding}. We also note that imaging in more bands with a catalogue overlay is available via the Aladin lite interface \citep{2022ASPC..532....7B} on the website hosting the catalogue\footnote{\url{http://astro.uni-tuebingen.de/~xrbcat}}.

Based on the literature search and inspection of the finding charts described above, we identified several problematic cases:

\begin{itemize}

    \item The X-ray source SAX~J0840.7$+$2248, which is classified as a LMXB in Liu07, is excluded from our catalogue. This transient is more likely to be the X-ray-rich gamma-ray burst GRB980429 rather than an XRB~\citep{2007ATel.1089....1S}.

    \item SWIFT~J0732.6$-$1330, which is listed as a LMXB in SIMBAD, is also not presented in our catalogue due to the fact that the source is an intermediate polar~(accreting magnetised white dwarf, IP)~\citep{2007A&A...475L..29B}. 

    \item  \citet{2021ApJ...916...80P}  find the LMXB candidate OGLE~BLG511.6~25872 to be of the same polar nature. We therefore do not include it in our catalogue.

    \item The X-ray transient Swift~J061223.0$+$701243(.9), listed as LMXB in Liu07, could also be an IP~\citep{2011A&A...526A..77B}. In this case, the authors suggest that an LMXB nature cannot be ruled out, and so the source can still be found in our catalogue; however, we urge caution when  considering the classification of this source.

    \item 2E~1613.5$-$5053~(listed as 1E~161348$-$5055.1 in Liu07 LMXB catalogue) is an X-ray source located close to the centre of the supernova remnant RCW~103. Based on XMM observations, it was discovered that the source has a 6.67 h periodicity, which may originate from either an orbital period of a LMXB or a spin period of an isolated neutron star~(INS)~\citep{2006Sci...313..814D}. After some time of searching for an optical/IR counterpart to confirm or refute a binary nature origin, \citet{2017ApJ...841...11T} finally reported the detection of an IR counterpart, the properties of which clearly indicate that the IR emission was coming from an INS, ruling out an LMXB origin scenario.

    \item V*~V934~Cen, another SIMBAD LMXB, seems to possess no XRB characteristics~(the corresponding link in SIMBAD for the suggested LMXB type does not provide any data regarding the source), and so it is excluded from our catalogue as well. 

    \item The same also goes for the binary TWA 22 (TWA 22AB), which is considered to be a LMXB in the  
    SIMBAD database. SIMBAD refers to \citet{2009A&A...506..799B} as the source of the corresponding classification. However, these authors estimated the total mass 
    of the binary to be about a $220 M_{\rm Jup}$, which is far too low for the XRB. Although the system 
    emits X-rays via coronal activity~\citep{2022A&A...661A..44S}, it cannot be classified as an XRB.

    \item 4U~1745$-$203 (H~1745$-$203) is a transient that was considered to be a separate LMXB, but the source appears to simply be a counterpart of SAX~J1748.9$-$2021 (NGC~6440~CX1), which has already been noted in Liu07.

    \item The same goes for bursting source MXB~1742$-$29. Following \citet{2007A&A...462.1065M}, we consider it to be the counterpart of LMXB 1A~1742$-$294.

    \item AX~J1620.1$-$5002 is also an X-ray transient, which for a long time was considered to be a separate XRB, but it is now strongly associated with the bursting transient LMXB MAXI~J1621$-$501~\citep{2018ATel11272....1C, 2018ATel11317....1G}. 

    \item 4U~0614$+$09 (4U~0614$+$091) is an UCXB with an uncertain orbital period, which is only constrained to be greater that 1\,h \citep{2014A&A...572A..99B}. However, following Liu07 and RK03 in the catalogue, we provide a value of 51.3\,min based on \citet{2008PASP..120..848S}.

    \item The nature of IGR~J17404$-$3655 is not well constrained, because it is unclear whether the source is a LMXB or a HMXB, and is potentially a BeXRB~\citep{2013A&A...560A.108C}. However, \citet{2018A&A...618A.150F} used K-band spectroscopy for the source, and revealed that the object is more likely to be a CV with a K3–5V donor star. Consequently, we made the decision to not include it.

    \item CXOGBS~J174623.5$-$310550 is an accreting binary, and is presented in our catalogue. However, its exact classification, that is, as either LMXB or CV, is still unknown~\citep{2019MNRAS.487.2296T}. 

    \item 1RXS~J180431.1$-$273932 was at first considered a symbiotic LMXB system~\citep{2007A&A...474L...1N} hosting a NS with a M5~III donor star. However, based on optical observations, \citet{2012A&A...544A.114M} excluded a possible symbiotic LMXB nature and identified the source as a magnetic CV. Therefore, despite the fact that 1RXS~J180431.1$-$273932 is listed as a symbiotic X-ray binary in \citet{2015AstL...41..114K}, we do not include it.

    \item Based on IR observations of 4U~1556$-$60 made by \citet{2013AstL...39..523R}, we associated an optical counterpart to this source, namely \textit{Gaia}~DR3 583304229928809907~(see Table~\ref{tabl:opt}), which lies well inside the position error circle of 0\farcs3. However, according to {the provided parallax}, the distance to this source should not exceed 1kpc~\citep{Gaiaedr3dist}, which is 3--4 times less than the distance estimations to 4U~1556$-$60~\citep{1997ApJS..109..177C, 2002A&A...391..923G}. We decided to leave this association with the \textit{Gaia}~DR3 source, but this discrepancy should be borne in mind. 

    \item For the source IGR~J17597$-$2201, \citet{2006A&A...453..133W} proposed 2MASS~J17594556$-$2201435 as a potential counterpart. However, based on Very Large Telescope (VLT) observations conducted by~\citet{2018A&A...618A.150F}, we can exclude this possibility.

    \item The persistent X-ray-bright LMXB 4U~1624$-$49 seems to have incorrect coordinates and optical/IR associations in the SIMBAD database. 
    The source was localised by \textit{Chandra} with 0\farcs6 uncertainty at the position
    $\alpha_{\rm J2000.0} = 16^{\rm h}28^{\rm m}02\fs825$, $\delta_{\rm J2000.0} = -49\degr11\arcmin54\farcs61$ \citep{2005ApJ...621..393W}, which corresponds to the known XMM source 2XMM~J162802.8$-$491154, which is located  $\approx 53\arcsec$ away from the position of 4U~1624$-$49 in SIMBAD. In our table, we provide the \textit{Chandra} coordinates and uncertainty, as well as the K-band data obtained by \citet{2005ApJ...621..393W}. In the corresponding \textit{Chandra} error circle, there are no \textit{Gaia}, 2MASS, or CatWISE2020 sources. We also label 2XMM~J162802.8$-$491154 as an alternative name for 4U~1624$-$49 due to the fact that this name is not listed as an identifier for this source in any other databases to the best of our knowledge.

    \item There is also an incorrect SIMBAD association with optical/IR archives for the LMXB GX~5$-$1 (4U~1758$-$25). At the position provided by SIMBAD, there is a bright star located $\approx 200$ pc away~\citep{Gaiadr3}, a distance which is totally inconsistent with distance estimations for GX~5$-$1~\citep[4.2--5.6 kpc, according to][]{2018ApJ...852..121C}. \citet{2000MNRAS.315L..57J} found an IR counterpart of the X-ray source at the position 
    $\alpha_{\rm J2000.0} = 18^{\rm h}01^{\rm m}08\fs222$, $\delta_{\rm J2000.0} = -25\degr04\arcmin42\farcs46$ with 0\farcs35 uncertainty, leading to it being located  $\approx 20\arcsec$ away from the position of GX~5$-$1 in SIMBAD. As in the case of 4U~1624$-$49, there are no \textit{Gaia}, 2MASS, or CatWISE2020  counterparts. However, in the error circle, we find a VVV source, for which the JHK magnitudes are provided in the table.

    \item We also decided to add the object named AX~J1659.0$-$4208 as a
    LMXB in our catalogue. The source was detected during the Advanced
    Satellite for Cosmology and Astrophysics (ASCA) Galactic plane
    survey~\citep{2001ApJS..134...77S}. At the time, the X-ray
    flux of the source was at the level of $6 \times 10^{-12}$ erg/s/cm$^{2}$
    ($0.7-10$\,keV), but its coordinates were not well constrained.
    Its accurate position was then derived by Chandra, which led
    to identification of a near-IR counterpart by
    \citet{2013AstL...39..523R}. These latter authors suggested that the source is
    likely to be a symbiotic LMXB or a CV. However, based on high
    extinction towards this area, and distance estimations of about 5-10
    kpc~\citep{2013AstL...39..523R}, unabsorbed X-ray luminosity can
    be estimated to be $10^{34}$  $\mathrm{erg}\,\mathrm{s}^{-1}$ at the very least (most likely even
    $10^{35}$$\mathrm{erg}\,\mathrm{s}^{-1}$), which is significantly
    greater than expected for a typical CV. We therefore suggest that
    the source is very likely to be a LMXB and it is added to our catalogue
    table.

    \item We add an X-ray source IGR~J19308$+$0530, which seems to be an IMXB hosting an NS and s F4V-type donor star~\citep{2018A&A...618A.150F}. At first it 
    was not clear whether the object is a CV or a L/IMXB~\citep{2013MNRAS.431L..10R}, and even the 
    SIMBAD database presents the source as a CV. However, based on the spectral analysis~\citep{2010MNRAS.408.1866R} and K-band 
    spectroscopy~\citep{2018A&A...618A.150F}, we suggest that IGR~J19308$+$0530 is most likely not a CV but an IMXB. It should also be noted that the star TYC~486$-$295$-$1 is a counterpart of IGR~J19308$+$0530~\citep{2008A&A...482..731R, 2010MNRAS.408.1866R}, but it is considered to be a separate source (also a CV type instead of XRB) by SIMBAD, $\approx 88\arcsec$ away from the INTEGRAL position for IGR~J19308$+$0530 listed in SIMBAD.
    
    \item  The source LMXB 4U~1728$-$34 was given an orbital period of $\sim 11$
    min~\citep{2010ApJ...724..417G}, and this value is also given for the binary in RK03. However, in an analysis of other X-ray observations, no further evidence was found for a periodic signal at 11 min. Subsequently, based on an analysis of its type-I X-ray burst, \citet{2020MNRAS.495L..37V} stated that the orbital period should be much higher, probably $\sim 66$ min or even $\gtrsim 2$ hr. In our catalogue, the value of 66 min is quoted, but readers should be aware of the lack of robust confirmation.

    \item The X-ray source J2039$-$5617 (listed as [SMD2015]~3 in SIMBAD) is presented in the RK03 catalogue as a LMXB. However, due to a likely association with the known RB source 1FGL~J2039.4$-$5621~\citep{2015ApJ...814...88S}, it is excluded from our catalogue.

    \item XB~1832$-$330 is a globular cluster LMXB (in NGC~6652) that was tentatively assigned a 43.6\,min orbital period~\citep{2000ApJ...530L..21D}. However, \citet{2012ApJ...747..119E} argue that the 43.6\,min candidate period is probably spurious and suggest a longer period (by a factor 3) of about 2.15\,hr (this value is quoted in our catalogue).

    \item There is uncertainty over the orbital period of the tMSP candidate 4FGL~J0540.0$-$7552. \citet{2021ApJ...917...69S} identified a consistent periodic signal at 2.7\,h in their photometry. However, this signal could represent either the orbital period, half the orbital period (if due to ellipsoidal variations), or could even be spurious. We decided to keep the 2.7\,h orbital period value in our catalogue table for this source, but other possibilities may be worth consideration.
    
 \end{itemize}

\begin{table}
\caption{\label{tabl:tMSPcands}tMSP candidates presented in our catalogue.}
\begin{tabular}{c c}
\hline\hline 
Source Name & References 
 \\ \hline

1FGL~J1417.7$-$4407 (J1417$-$4402 in RK03) & [1] \\ 
CXOGlb~J183544.5$-$325939 (NGC~6652B) & [2] \\ 

4FGL~J0427.8$-$6704 & [3], [4] \\ 

4FGL~J0407.7$-$5702 & [5] \\ 

CXOGlb~J174804.5$-$244641 (Terzan~5~CX1) & [6] \\

CXOU~J110926.4$-$650224 & [7], [8] \\

3FGL~J1544.6$-$1125 & [9], [10], [11] \\
 
4FGL~J0540.0$-$7552 & [12] \\

RX~J1739.4$-$2942 (GRS~1736$-$297)  & [13] \\

\hline
\end{tabular}

\vspace{0.2cm}
\textbf{Notes.}
Along with the candidates, all three confirmed tMSPs: XSS~J12270$-$4859, IGR~J18245$-$2452 and PSR~J1023$+$0038 are also included in the catalogue.
\\
\textbf{References.}
[1] \citet{2018ApJ...866...83S}, 
[2] \citet{2021MNRAS.506.4107P},  
[3] \citet{2020MNRAS.494.3912K}, 
[4] \citet{2020ApJ...895...89L},
[5] \citet{2020ApJ...904...49M},
[6] \citet{2018ApJ...864...28B},
[7] \citet{2019A&A...622A.211C},
[8] \citet{2021A&A...655A..52C},
[9] \citet{2015ApJ...803L..27B},
[10] \citet{2017ApJ...849...21B},
[11] \citet{2021ApJ...923....3J},
[12] \citet{2021ApJ...917...69S}, 
[13] \citet{2016ATel.8744....1T}.
\end{table}

As mentioned in Sect. 1, we decided to add all three known tMSP: XSS~J12270$-$4859, IGR~J18245$-$2452, and PSR~J1023$+$0038 (all three are also listed in RK03 by the names J1227$-$4853, J1824$-$2452, and AY~Sex, respectively). In addition, we include some tMSP candidates identified as such in the literature, which can be found in Table~\ref{tabl:tMSPcands} with references. In the catalogue table itself, one can find a corresponding flag column named `tMSP\_Flag'~(see Table~\ref{tabl:cols}).

Several objects originally listed in the HMXB catalogue of \citet{Liu_HMXB} are now moved to the current LMXB catalogue as their preferred classification changed:

\begin{itemize}
\item 
According to \citet{Karasev08}, XTE~J1901$+$014 was suggested to be the first low-mass fast X-ray transient. However, the exact nature of this source remains unclear. Nevertheless, according to \citet{Sato19}, it appears to be more similar to a LMXB, hence its inclusion in our catalogue.

\item 
IGR~J16358$-$4726 was initially identified as an HMXB by \citet{Chaty08}. However, \citet{Nespoli10} later invalidated this identification, and now the source is classified as a symbiotic X-ray binary, which we consider to be a LMXB subclass based on the typical donor mass.

\item 
The situation is similar for IGR~J17407$-$2808. This source was first classified as part of a supergiant fast X-ray transients~\citep[SFXT,][]{2006ApJ...646..452S}, which are all HMXBs. However, subsequent works~\citep{2009ApJ...701.1627H, 2011ATel.3688....1G} ruled out the presence of a supergiant companion, and showed that this source is  more likely to
be a LMXB with an F-type dwarf. \citet{2011ATel.3695....1K} also found that a few days after the outburst, the IR counterpart was 1 mag brighter. Such behaviour is more expected during the LMXB outburst due to the exposure of the optical component to the X-rays emitted by the compact object.

\item 
After analysing the X-ray spectra and near-IR observations of SAX~J1452.8$-$5949, \citet{Kaur09} were able to estimate the distance to SAX~J1452.8$-$5949 ($\lesssim 10$ kpc). This estimation eliminated the possibility of the system being a HMXB, as it would have to be an extragalactic source in that case. Consequently, the authors suggested that the binary has a low-mass companion and could either
be a LMXB or an IP. Therefore, we classify SAX~J1452.8$-$5949 as a LMXB candidate.

\item 
It has been unclear for some time whether 4U~1807$-$10 is a HMXB or a LMXB. However, based on the ratio of spin and orbital periods \citep{Blay08}, as well as the fact that the system shows type-I X-ray bursts \citep{2017AstL...43..781C}, we conclude that an LMXB origin is more likely for this source, despite the fact that 4U~1807$-$10 is marked as a HMXB candidate in \citet{Blay08}, and therefore we add 4U~1807$-$10 to our LMXB catalogue.

\end{itemize}

In addition, there is one source that is present in both of our catalogues, in this one, and in the one dedicated to HMXBs~\citep{Marvin23}. This is due to the presence of conflicting results and the absence of any final decision on the true nature of the source:

\begin{itemize}

\item 
Cir~X$-$1 was originally classified as a LMXB after its discovery. However, \citet{2013ApJ...779..171H} were able to determine that the system is only about 4500 years old, which led to a revision of its classification. Additionally, \citet{2007MNRAS.374..999J} reported that Cir~X$-$1 contains a supergiant companion of A0 to B5 type. 
However, Cir~X$-$1 has also exhibited type-I X-ray bursts~\citep{1986MNRAS.221P..27T}, which are characteristic of LMXBs. Furthermore, the fact that the companion star in Cir~X$-$1 has yet to be decisively detected at optical wavelengths supports a possible LMXB nature for the system. \citet{2016MNRAS.456..347J} concluded that the optical counterpart in Cir~X$-$1 could either be an unevolved low-mass star or a giant star that is still in the process of recovering from the supernova explosion that occurred less than 5000 years ago. Due to the ambiguity surrounding the nature of Cir~X$-$1, the source has been included in both our LMXB and HMXB catalogues.
\end{itemize}

As mentioned previously, several objects included in RK03 and Liu07 have been removed from our catalogue. Sometimes this was necessary because objects reported as independent in RK03 and Liu07 represented the same source (in either of the catalogues). This was mainly due to large positional uncertainties at the time of the catalogue compilation, or in some cases because the objects are extragalactic. In the case of the former, we include only the element from the pair in our LMXB list with the most used name, omitting the other, and summing up all known data. In the extragalactic origin scenario, the corresponding source is removed as we only consider Galactic LMXBs. Our initial source list was therefore modified as follows:

\begin{itemize}

    \item  The relatively bright source 2A~0521$-$720, which is more commonly referred to as LMC~X$-$2, is located in the LMC (as the name suggests). As we consider only Galactic objects, we exclude it from our catalogue.
    
    \item  The same is true for RX~J0532.7$-$6926 (located in LMC), and so it is also not included in the current catalogue.
    
    \item  Two sources AX~J1745.6$-$2901 and 1A~1742$-$289 listed in Liu07 actually correspond to the same LMXB, and as the name `AX~J1745.6$-$2901' is more commonly used, we combine information regarding the two from Liu07 under this name.
        
    \item The second case of identical sources is 1E~1743.1$-$2852 and SAX~J1747.0$-$2853, and so again we gave the most popular name to the object, SAX~J1747.0$-$2853.
    
 \end{itemize}

 \subsection{Summary of other differences with respect to the Liu07 and RK03 catalogues}

In addition to the problematic cases described above, and the overall expansion of the sample, several other changes have been made compared to Liu07 and RK03:

 \begin{itemize}   

    \item To standardise the reported X-ray fluxes, we revised the energy ranges used in the Liu07 study. Instead of using a single range of 2--10 keV, we now present `soft' and `hard' X-ray fluxes separately, each in well-defined energy ranges.
    
    \item In addition, X-ray fluxes (absorbed) in our table are represented in cgs units ($\mathrm{erg}\,\mathrm{s}^{-1}\,\mathrm{cm}^{-2}$, instead of Jy reported in Liu07), which are more commonly used and meaningful for X-ray sources. We also now report both maximum and minimum flux values for each of the two energy bands instead of only reporting the maximum, as done in Liu07.

    \item To better reflect various phenomenological features reported in the literature, we increased the number of flags used to describe the properties of the sources. In comparison to Liu07, where 11 flags were used, we use 17.

    \item Our new version of the LMXB catalogue has additional~(in comparison with Liu07) columns for hydrogen column density, near- and mid-IR data (from 2MASS, VVV, WISE, literature, etc.), BP-RP colour, optical \textit{Gaia}/StarHorse/literature data, and distances.

    \item In addition, we include up to six finding charts (from mid-IR images to hard X-ray bands) for each object, instead of just a reference to a finding chart.

    \item Finally, our catalogue now includes 349 objects, which is a 71\% increase from the 204 objects listed in Liu07 and RK03 combined.

\end{itemize}

\section{Summary and conclusions}\label{conc}
In this paper, we present a new LMXB catalogue, which provides an up-to-date census of known LMXBs in the Galaxy with a variety of corresponding multi-wavelength information extracted from current optical/IR surveys. In total, the catalogue contains 349 sources and has 62 columns. Descriptions of the columns are presented in Table~\ref{tabl:cols}. The full version of the catalogue can be obtained through the VizieR database\footnote{\url{https://cdsarc.cds.unistra.fr/viz-bin/cat/J/A+A/675/A199}}, and can also be found on a dedicated website\footnote{\url{http://astro.uni-tuebingen.de/~xrbcat}}. We ask authors who use our catalogue in their research to refer to us by citing this article, as well as by referring to the website link in either the acknowledgements or a footnote.

We would like to emphasise here that the origin of some of the sources listed in the catalogue is not yet certain; some sources have only been tentatively classified as LMXBs due to the similarity of their X-ray properties to those of the systems identified, with no counterparts found in other bands.
Along with this article, a new similar catalogue of HMXBs is being published \citet{Marvin23}. We hope that both catalogues will be useful tools for future studies, which includes (but is not limited to) population studies of various types of X-ray-emitting objects (especially XRBs), searches for new XRBs in ongoing and upcoming surveys, classification of unidentified X-ray sources, and as training data-sets for machine learning classification algorithms~\citep{Yang2021, Yang2022}. In particular, our catalogue be relevant in the context of the ongoing analysis of eROSITA survey data, which is expected to substantially increase the number of known XRBs \citep{2014A&A...567A...7D}. We therefore plan to update our catalogues to the best of our ability to keep the ever-increasing census of the Galactic LMXBs up to date and accessible to the community.

\begin{acknowledgements}
  This research has made use of the SIMBAD data base and
  VizieR catalogue access tool operated at CDS, Strasbourg,
  France, and NASA’s Astrophysics Data System (ADS). This research has
  made use of "Aladin sky atlas" developed at CDS, Strasbourg
  Observatory, France. This work has made use of data from the
  European Space Agency (ESA) mission \textit{Gaia}
  (\url{https://www.cosmos.esa.int/gaia}), processed by the
  \textit{Gaia} Data Processing and Analysis Consortium (DPAC,
  \url{https://www.cosmos.esa.int/web/gaia/dpac/consortium}). Funding
  for the DPAC has been provided by national institutions, in
  particular the institutions participating in the \textit{Gaia}
  Multilateral Agreement. We acknowledge the public data from
  \textit{XMM-Newton}, CXO, ROSAT, \textit{Swift},
  INTEGRAL, WISE, 2MASS and VISTA.
  AA thanks Deutsche Forschungsgemeinschaft (DFG) for support through
  the eRO-STEP research unit project 414059771 (DO 2307/2-1 and WI
  1860/19-1).
\end{acknowledgements}

\bibliographystyle{aa_url}
\bibliography{lmxbmain}

\clearpage
\onecolumn

\begin{appendix}

\section{Optical/IR counterparts}\label{append1}

\begin{longtable}
{p{4.9cm}p{4.0cm}p{4.0cm}p{4.0cm}}
\caption{\label{tabl:opt} Identified archive optical/IR counterparts of LMXBs.}\\
\hline\hline

Source name & \textit{Gaia}~DR3 & 2MASS & CatWISE2020 \\
\hline
\\
\multicolumn{4}{c}{Based on optical/IR/radio follow-ups:}
\\
\\
\endfirsthead
\multicolumn{4}{c}
{{\bfseries \tablename\ \thetable{}. continued}} \\
\hline\hline
Source name & \textit{Gaia}~DR3 & 2MASS & CatWISE2020\\ 
\hline
\endhead

\hline
\endlastfoot

MAXI~J0556$-$332 & 2890346074897001216 &  &  \\

4FGL~J0427.8$-$6704 &
4656677385699742208 
&
J04274958$-$6704350 
&
J042749.64$-$670435.0 \\

IGR~J17494$-$3030 &
4056028099172996352 &  &  \\

1FGL J1417.7-4407 &
6096705840454620800 
&
J14173057$-$4402574 
&
J141730.57$-$440257.5 \\

XTE~J1637$-$498 &
5940421734427982336 
&
J16370267$-$4951401 
& \\

MAXI~J1807$+$132
&
4497207964419829632 &  &  \\

3A~1837$+$049 &
4283919201304278912 
&
J18395759$+$0502113 
&
J183957.57$+$050210.5 \\

MAXI~J1828$-$249 &
4076998775244833280 &  &  \\

1RXS~J180408.9$-$342058  &
4042163562572848384 &  &  \\

IGR~J17091-3624 &
5976921951382731520 
& 
& 
J170907.75$-$362425.0 \\

IGR~J17585$-$3057 &
4044163406621956992 
& 
&
J175829.80$-$305702.3 \\

4U~1543$-$624 &
5826501373348972288  &  &  \\

Swift~J1858.6$-$0814 &
4203799300867262720 &  &  \\

MAXI~J1305$-$704 &
5843823766718594560 
&  
& 
J130655.74$-$702704.2 \\

IGR~J17329$-$2731  &
4061336747511224704 
&
J17325067$-$2730015
&
J173250.54$-$273003.4 \\

IGR~J16358$-$4726 
&
&
J16355369$-$4725398
&
J163553.75$-$472541.0 \\

XMMU~J174445.5$-$295044 
&
&  
J17444541$-$2950446
&
J174445.43$-$295044.5 \\

IGR~J17454$-$2919 &
& 
J17452768$-$2919534
&  \\

4U~1705$-$250  &
4112450294268643456  &  &  \\

4FGL~J0540.0$-$7552 &
4648562676357022208 
&
&  
J054001.80$-$755419.7 \\

3FGL~J1544.6$-$1125  &
6268529198286308224 
&
&  
J154439.38$-$112804.5 \\

CXOU~J110926.4$-$650224  &
5240167590731178624
&
& 
J110926.21$-$650227.1 \\

4FGL~J0407.7$-$5702 &
4682464743003293312 
&
&  
J040731.65$-$570025.1 \\

1RXH~J173523.7$-$354013 
&
5974787971132982144
&
&  
 \\
  
IGR~J16287$-$5021
&
5934583950467938304
&
&  
 \\

4U~1556$-$60
&
5833042299288099072
&
&  
 \\

EXO~1846$-$031
&
4258794192383316864
&
J18491710$-$0303559
&  
J184917.08$-$030355.3
 \\

AX~J1735.8$-$3207
&
4055019091071763712
&
J17354627$-$3207099
&  
 \\

IGR~J16293$-$4603
&
5942210433690960640
&
J16291285$-$4602506     
&  
J162912.85$-$460250.4   
\\

4U~1626$-$67
&
5809528276749789312 
&
&  
\\

SAX~J1810.8$-$2609
&
4064636102946481408 
&
&  
\\

XTE~J1709$-$267
&
4108865783346597760
&
&  
\\

XTE~J1710$-$281
&
4108449755564592256
&
&  
\\

IGR~J19308$+$0530
&
4294249387962232576     
&
J19305075$+$0530582     
& 
J193050.75+053058.0     
\\

Swift~J2037.2$+$4151
&
&
J20370560$+$4150051
&  
J203705.60$+$415005.2 
\\

GS~1826$-$238
&
4077372536242802944
&
&
\\

SRGA~J181414.6$-$225604
&
4066460746608630912
&
J18141475$-$2256195
&
J181414.78$-$225619.4
\\
\\
\\

\multicolumn{4}{c}{Based on \citet{Arnason21}:}
\\
\\

GRO~J1655$-$40 & 
5969790961312131456 
& 
J16540014$-$3950447 
& 
J165400.13$-$395044.7 \\

SWIFT~J061223.0$+$701243 & 1107229825742589696 
&  
& J061222.63$+$701243.1 \\

1A~1246$-$588 & 
6059778089610749440 
&
&  
J124939.17$-$590516.3 \\

MXB~1659$-$29 & 
6029391608332996224  &  &  \\

SAX~J1711.6$-$3808 & 
5973177495780065664 
& 
J17113714$-$3807073 
& 
J171137.14$-$380707.4 \\

4U~1724$-$307 & 
4058208396397618688 
 & 
J17273315$-$3048076 
 & 
J172733.28$-$304809.0 \\

EXO~1747$-$214 & 
4118590585673834624  &  &  \\

4U~1755$-$33 & 
4042473487415175168  &  &  \\

HETE~J1900.1$-$2455 & 
4074363039644919936 &  &  \\

4U~1915$-$05 & 
4211396994895217152  &  &  \\

XTE~J1901$+$014 & 
4268294763113217152 &  &  \\

\end{longtable}

\clearpage

\section{Catalogue format}\label{append2}

\begin{longtable}{p{0.2cm}p{2.6cm}p{1.9cm}p{12.0cm}}
\caption{\label{tabl:cols} Description of columns in the LMXB catalogue table.}\\
\hline\hline
№ & Column name & Unit & Description \\
\hline
\endfirsthead
\multicolumn{4}{c}
{{\bfseries \tablename\ \thetable{}. continued}} \\
\hline\hline
№ & Column name & Unit & Description \\ 
\hline
\endhead

\hline
\endlastfoot
    1 & 'Name' &  &  Object name, which is the most common name in the literature.
    \\
    2 & 'RAdeg' &  deg  & Right Ascension in degrees (ICRS). 
     \\
    3 & 'DEdeg' &  deg &  Declination in degrees (ICRS).
    \\
    4 &  'PosErr' &  arcsec & Positional error in arcseconds. 
    \\
    5 &  'Coord\_Ref' &  &  The source from which the position of the object is taken. 
    \\
    6 &  'GLON' & deg & Galactic longitude in degrees. 
    \\
    7 & 'GLAT' & deg &  Galactic latitude in degrees. 
    \\
    8 & 'Xray\_Type' & &  List of the X-ray types assigned to the object. For more information see Sect.~\ref{sec:xtypes}.
    \\
    9 &  'Porb' & day &  Orbital period of the binary system in days.
    \\
    10 &  'Ppulse' & s &  Pulsation period (spin) of a NS in seconds.
    \\
    11 &  'Alt\_Name' & &  Most frequently used alternative name.
    \\
    12 &  'SpType' & &  Spectral type of the optical counterpart.
    \\
    13 & 'Vmag' & mag &  Optical magnitude in V-band according to SIMBAD database and literature. 
    \\   
    14 & 'e\_Vmag' & mag &  Corresponding  magnitude error in V-band.
    \\
    15 &  'Gaia\_DR3\_ID' & &  Name in Gaia DR3(eDR3) catalogue.
    \\    
    16 &  'Gmag' & mag &  Optical magnitude in G-band according to a Gaia catalogue or the literature. 
    \\     
    17 &  'e\_Gmag' & mag &  Corresponding  magnitude error in G-band.     
    \\     
    18 & 'BP-RP' & mag &  BP-RP colour. 
    \\    
    19 & '2MASS\_ID' &  &  Name in 2MASS catalogue.
    \\  
    20 & 'Jmag' & mag &  IR magnitude in J-band according to 2MASS or the literature.
    \\     
    21 & 'e\_Jmag' & mag &  Corresponding  magnitude error in J-band according to 2MASS or the literature. 
    \\
    22 &  'Hmag' & mag &  IR magnitude in H-band according to 2MASS or the literature.
    \\
    23 & 'e\_Hmag' & mag &  Corresponding  magnitude error in H-band according to 2MASS or the literature.  
    \\
    24 & 'Kmag' & mag &  IR magnitude in K-band according to 2MASS or literature.
    \\
    25 & 'e\_Kmag' & mag &  Corresponding  magnitude error in K-band according to 2MASS or the literature.
    \\
    26 & 'WISE\_ID' &  &  Name in CatWISE, AllWISE, or WISE catalogue.
    \\  
    27 & 'W1mag' & mag &  IR magnitude in W1 WISE band.  
    \\   
    28 & 'e\_W1mag' & mag &  Corresponding magnitude error in W1 WISE band.
    \\
    29 & 'W2mag' & mag &  IR magnitude in W2 WISE band.   
    \\   
    30 & 'e\_W2mag' & mag &  Corresponding magnitude error in W2 WISE band.
    \\     
    31 & 'N\_H' & 10$^{22}$$\mathrm{cm^{-2}}$ & Neutral hydrogen column density.
    \\   
    32 & 'MinSoftFlux' & $10^{-12}$erg/s/cm$^{2}$  &  Minimum flux (absorbed) from a source in soft X-ray.
    \\ 
    33 & 'MaxSoftFlux' & $10^{-12}$erg/s/cm$^{2}$  &  Maximum flux (absorbed) from a source in soft X-ray.
    \\ 
    34 & 'Soft\_Xray\_Var& & Ratio of maximal soft X-ray flux and minimal soft X-ray flux.
    \\
    35 & 'Soft\_Xray\_Range' &keV & Soft X-ray flux energy range used.
    \\
    36 & 'MinHardFlux' & $10^{-12}$erg/s/cm$^{2}$ & Minimum flux (absorbed) from a source in hard X-ray.
    \\
    37 & 'MaxHardFlux' & $10^{-12}$erg/s/cm$^{2}$ &  Maximum flux (absorbed) from a source in hard X-ray. 
    \\
    38 & 'Hard\_Xray\_Var' & & Ratio of maximal hard X-ray flux and minimal hard X-ray flux.
    \\
    39 & 'Hard\_Xray\_Range' & keV & Hard X-ray flux energy range used.
    \\
    40 & 'AG' & mag  & G-band absorption $A_{\rm G}$.
    \\ 
    41 & 'Mean\_Dist' & pc &  Mean estimate of the distance to the binary system.
    \\
    42 & 'Low\_Dist' & pc &  The lowest (minimum) estimate of the distance to the binary system. 
    \\     
    43 & 'High\_Dist' & pc &  The highest (maximum) estimate of the distance to the binary system. 
    \\    
    44 & 'Incl' & deg & Binary system inclination.
    \\ 
    45 & 'Mx' & $\mathrm{M_{\odot}}$ &  Mass of the compact object.
    \\     
    46 & 'Mopt' & $\mathrm{M_{\odot}}$ &  Mass of the optical counterpart. 
    \\
    47 &'B-V' & mag &  B-V colour index.
    \\     
    48 & 'E(B-V)' & mag &  E(B-V) colour excess.
    \\     
    49 & 'tMSP\_Flag' &  &  tMSP flag, which can take one of the two values, indicating that the source is:
    \\ & & & 1 --- one of the three solidly considered tMSPs.
    \\ & & & 2 --- a tMSP candidate.
    \\    
    50 & 'Liu07\_IDX' & &  Ordinal number of the source in the table of Liu07 LMXB catalogue.
    \\     
    51 & 'RK03\_IDX' & &  Ordinal number of the source in the table of RK03 LMXB catalogue.   
    \\
    52 & 'Orb\_Ref'& & Orbital period origin reference.
    \\
    53 & 'Pulse\_Ref'& & Spin (pulsation) period of the NS origin reference. 
    \\
    54 & 'Spect\_Ref' & & Reference to the  origin of the optical  spectral type of the  star.
    \\
    55 & 'Vmag\_Ref'& & Reference to the origin  of the V-band magnitude.   
    \\
    56 & 'Gmag\_Ref' & & Reference to the origin of the G-band magnitude. 
    \\
    57 & 'JHK\_Ref' & & Reference to the origin  of the JHK information. 
    \\
    58 & 'WISE\_Ref' &  &  Reference to the WISE IR catalogue used (CatWISE, AllWISE or WISE).
    \\ 
    59 & 'N\_H\_Ref' & & $N_{\rm H}$ value origin reference. 
    \\
    60 & 'AG\_Ref' &   & Reference to the origin of the G-band absorption $A_{\rm G}$ .
    \\
    61 & 'Dist\_Ref' & & Reference to the origin of the  distance estimations. 
    \\
    62 & 'Mass\_Ref' & & Reference to the origin of the  mass estimations. 
\end{longtable}
%\normalsize

\end{appendix}

%%%%%%%%%%%%%%%%%%%%%%%%%%%%%%%%%%%%%%%%%%%%%%%%%%

% Don't change these lines
% typesetting comment
%\label{lastpage}

\end{document}